\documentclass[aps,pra,twocolumn,showpacs,superscriptaddress,tightenlines,preprintnumbers,sort&compass,amsmath,amssymb,floatfix,nofootinbib]{revtex4-1}

\usepackage{hyperref}
\usepackage{amsmath}
\usepackage{graphicx}
\usepackage{amssymb}
\usepackage{epsfig}
\usepackage{amsfonts}
\usepackage{color}

\usepackage[utf8x]{inputenc}
\usepackage[T1]{fontenc}

\newcommand{\etal}{{\em et al.}}

\def \>{\rangle}

\newcommand{\ket}[1]{\mbox{$|#1\rangle$}}

\newcommand{\pla}{Phys. Lett. A~}

\def\yyy#1{#1}

\def\extra#1{{``#1''}}

\begin{document}

\title{\yyy{Direct method for measuring} and witnessing quantum entanglement of arbitrary two-qubit states \yyy{through
Hong-Ou-Mandel interference}}

\author{Karol Bartkiewicz}
\email{bark@amu.edu.pl} \affiliation{Faculty of Physics, Adam
Mickiewicz University, PL-61-614 Pozna\'n, Poland}
\affiliation{RCPTM, Joint Laboratory of Optics of Palack\'y
University and Institute of Physics of Academy of Sciences of the
Czech Republic, 17. listopadu 12, 772 07 Olomouc, Czech Republic }

\author{Grzegorz Chimczak}
\affiliation{Faculty of Physics, Adam Mickiewicz University,
PL-61-614 Pozna\'n, Poland}

\author{Karel Lemr}
\affiliation{RCPTM, Joint Laboratory of Optics of Palack\'y
University and Institute of Physics of Academy of Sciences of the
Czech Republic, 17. listopadu 12, 772 07 Olomouc, Czech Republic }

\begin{abstract}
We describe a direct method for experimental determination of
the negativity of an arbitrary two-qubit state
with 11 measurements performed
on  multiple copies of the two-qubit system.
\yyy{Our method is based on the experimentally accessible sequences
of singlet projections performed on up to four qubit pairs.} 
In particular, our method permits \yyy{the} application 
of the Peres-Horodecki separability criterion
to an arbitrary two-qubit state. 
We explicitly demonstrate that 
measuring entanglement in terms of  negativity requires 
three  measurements more than detecting two-qubit entanglement.
The reported minimal set of \yyy{interferometric} 
measurements provides a complete description of 
bipartite quantum entanglement in 
terms of two-photon interference.
This set is smaller than the set of 15 measurements 
needed to perform a complete quantum state tomography
of an arbitrary two-qubit system. \yyy{Finally, we demonstrate that the
set of 9 Makhlin's invariants needed to express the negativity
can be measured by performing 13 multicopy projections.
We demonstrate that these invariants are both a useful theoretical concept 
for designing specialized quantum interferometers and that
their direct measurement within the framework of linear optics 
does not require performing complete quantum state tomography.} 
\end{abstract}



\date{\today}
\maketitle

\section{Introduction}

Local invariants describe the nonlocal properties 
of quantum systems and can be applied to check if 
two quantum systems are locally equivalent~\cite{Grassl1998}, 
i.e., if they can be transformed into one another only via local 
unitary operations on their subsystems.
Over the last years, it was shown that local 
invariants of quantum systems are very useful 
in quantum information processing. In particular,
it was also shown that the invariants of quantum codes can be a
useful tool in quantum error correction~\cite{Ranis2000}
necessary for advanced quantum computations or simulations. 
Moreover, for the two-qubit case, Makhlin~\cite{Makhlin00} showed
that 18 invariants can be used to characterize two-qubit 
gates (see also~Ref.~\cite{Koponen2005}) and arbitrary two-qubit states. 
The two-qubit case is the most interesting 
for practical applications such as  quantum communications~\cite{Riedmatten04}
and quantum cryptography~\cite{Ekert91}. 
Two-qubit invariants were also analyzed by King and Welsh
in Ref.~\cite{King06}. The authors found 21 fundamental
invariants of a two-qubit state. Recently, the local unitary invariants of  
multi-qubit states have been described by 
Jing {\it et al.} in Ref.~\cite{Jing2015}. These authors
demonstrated that some of the formerly 
studied two-qubit invariants are algebraically dependent
and they provided a set of 12 independent invariants
for two-qubit states.

One of the natural applications of local invariants
is  detecting  and quantifying 
quantum entanglement~\cite{Schrodinger35,Einstein35}.
In particular, they can be used to measure entanglement 
monotones~\cite{Osterloh12}. 
It was demonstrated by Carteret~\cite{Carteret05} 
that the two-qubit invariants of Kempe~\cite{Kempe1999} can
be applied to design quantum circuits for detecting 
quantum entanglement via the Peres-Horodecki 
criterion~\cite{Peres96,Horodecki96}. 
A more detailed analysis of  this problem was performed by
Bartkiewicz {\it et. al} in Refs.~\cite{Bart15a,Bart15b}.
In particular in Ref.~\cite{Bart15b} it was explicitly 
shown that 9 of 18 Makhlin's invariants can be used
to measure the negativity~\cite{Zyczkowski98,Vidal02} 
of an arbitrary two-qubit quantum state.
This negativity is directly related to the logarithmic negativity,
which is an entanglement measure with a clear physical interpretation.
Partial results for expressing concurrence~\cite{Wootters98},
an alternative entanglement measure related to the entanglement
of formation, via local invariants were reported in 
Ref.~\cite{Chaudhary2016,Carteret06}. 
For a restricted class of states the concurrence
was measured in a simple experimental setup~ \cite{Walborn06}. 
Many other interesting results on measuring the concurrence
were reported also in Refs.~\cite{Aolita06, Zhou14}.
For comparison of  negativity and concurrence
as two-qubit entanglement measures see Ref.~\cite{Verstraete01,Miran04}.
The whole topic of quantum entanglement 
was also reviewed in several works, e.g., Refs.~\cite{BengtssonBook,Horodecki09,SchleichBook}.

Despite these many interesting  results there are still
some open problems regarding direct experimental
detection and quantification of quantum entanglement~\cite{WernerList,Guhne07,Mintert07,Guhne09}. 
This might be due to the fact that measuring entanglement
even in the bipartite case is NP-hard problem~\cite{Gurvits2003,Gharibian2010}
and it cannot be performed with a single copy of
a given bipartite state without full quantum state tomography~\cite{Lu2016}.
In this paper we will demonstrate how to solve this problem
for a general two-qubit case and the negativity 
as an entanglement measure.

The problem of measuring negativity approximately
was initially studied in Refs.~\cite{Horodecki02,Horodecki03}.
In this paper, we express the 9 relevant local invariants
of Makhlin in terms of 13 more fundamental quantities 
that are measurable directly with interferometers.
By applying our approach one can
measure the negativity of an arbitrary two-qubit state
by measuring 11 parameters
or detect entanglement in any two-qubit state
by measuring 8 parameters 
with simpler setups than initially proposed 
in Refs.~\cite{Carteret05,Augusiak08,Bart15a}.
The most popular way to measure
the entanglement of a given state $\hat \rho$ is to reconstruct
this state by measuring at least 15 parameters,
and to calculate any entanglement measures for $\hat \rho$.
However, in this way we also acquire some unnecessary information
related to local properties of $\hat \rho$ (see, e.g., Ref.~\cite{Maciel09}) 
With deterministic sources of two-qubit states
and highly efficient detectors, the presented approach 
could be more efficient than quantum state tomography.

Here,  we present the first experimentally-feasible scheme 
for detecting and measuring quantum entanglement
of a given two-qubit state.
To detect entanglement we apply the Peres-Horodecki
separability criterion~\cite{Peres96,Horodecki96} 
given in terms of the sign of determinant of a given 
partially-transposed two-qubit 
density matrix \cite{Augusiak08,Demianowicz11}. 
There are other methods of detecting entanglement, including
the adaptive method of Park \etal~\cite{Park10},
measuring fully-entangled fraction~\cite{Bartkiewicz16fef}
which detects the entanglement of all entangled
Werner states; the collective witness of 
Rudnicki~\etal~\cite{Rudnicki11,Lemr16}; 
or the entropic entanglement witness investigated
in Ref.~\cite{Bovino05}.
However, the determinant of a partially-transposed
density matrix detects the quantum entanglement 
of all two-qubit entangled state.
Moreover, it is especially well suited to be 
studied in terms of local invariants and 
their interferometric constituents. 
Our analysis reveals a fundamental
difference in detecting and quantifying quantum entanglement.
This difference was not apparent as both the two-qubit
negativity and universal entanglement witness
were analyzed as functions of the same moments 
of a given partially-transposed density 
matrix~\cite{Bart15a,Bart15b}.

This paper is organized as follows: in Sec.~II,  negativity is
defined as a function of the relevant Maklin's invariants;
in Sec.~III, these invariants are defined via 
experimentally-accessible state projections on multiple 
copies of the two-qubit state. In particular, we show
that one needs the same information to measure the
values of the relevant Makhlin's invariants and
to determine the negativity.
In Sec.~IV we describe a direct method for
measuring the multicopy projections with
linear optics.
Next, we discuss the operational 
difference between measuring and detecting quantum 
entanglement within our framework. We conclude in Sec.~V.

\section{Theoretical framework}

Negativity is an important entanglement measure with
a clear operational meaning as the entanglement cost
under operations preserving the positivity of partial
transpose (PPT)~\cite{Audenaert03,Ishizaka04}. 
Other interpretations relate negativity to the number 
of dimensions of two entangled subsystems~\cite{Eltschka13}. 
Formally, it is defined as a quantitative
version of the Peres-Horodecki separability criterion~\cite{Peres96,Horodecki96}.
It was first introduced by \.Zyczkowski \etal~\cite{Zyczkowski98}
and subsequently proved to be an entanglement measure
by Vidal and Werner~\cite{Vidal02}.
In particular, for two-qubit density matrices $\hat \rho$, it can be defined 
as the only positive solution (see Ref.~\cite{Rana13}) of 
the following equation for $N$ 
\cite{Bart15b}
\begin{equation}\label{eq:neg}
a_4N^4 + a_3N^3 +a_2N^2 +a_1N+a_0 =0,
\end{equation}
where $a_0=48\det\hat \rho^\Gamma,$ $a_1=4(1-3\Pi_2+2\Pi_3),$ $a_2=6(1-\Pi_2),$
$a_3=6,$ $a_4=3,$ and
the moments of the partially-transposed
density matrix $\hat\rho^\Gamma$ are given as $\Pi_n=\mathrm{tr}[(\hat \rho^{\Gamma})^n]$.
In our definition of  two-qubit negativity $N=2\mu$
where $\mu$ is the absolute value of the negative eigenvalue of
$\hat\rho^\Gamma.$
Interestingly, solving Eq.~(\ref{eq:neg}) was shown to provide simpler formulas
for negativity than other equivalent approaches~\cite{Miran15}.
The determinant of the partially-transposed density matrix can be expressed
as \cite{Augusiak08}
\begin{equation}\label{eq:drg}
\det\hat\rho^\Gamma = \tfrac{1}{24}(1-6\Pi_4+8\Pi_3+3\Pi_2^2-6\Pi_2).
\end{equation}
By studying the sign of this determinant one can detect 
the entanglement for an arbitrary two-qubit state.
If there is no negative solution, the negativity equals zero.
In Ref.~\cite{Bart15a} it was shown that 
the moments of the partially-transposed density matrix
are given as
\begin{eqnarray}
4\Pi_2&=&1+x_1\nonumber\\
16\Pi_3&=&1+3x_1+6x_2\\
64\Pi_4&=&1+6x_1+24x_2+x_1^2+2x_3,\nonumber
\end{eqnarray}
where
$x_1=I_2+I_4+I_7$, 
$x_2=I_1+I_{12}$, 
$x_3=I_2^2-I_3+2(I_5+I_8+I_{14}+I_4 I_7)$
are defined in terms of Makhlin's invariants 
$I_n$ for $n=1,2,3,4,5,7,8,12,14.$ 
\yyy{From Refs.~\cite{Bart15a,Bart15b} it }could appear that we need the same amount
of experimental data to determine both $\det\hat\rho^\Gamma$
and negativity $N$. However, this is not
the case as we will demonstrate in the following sections.
The 18 invariants described by Makhlin in Ref.~\cite{Makhlin00}
are expressed in terms of the correlation matrix 
$\hat{\beta}$ 
with elements $\beta_{ij}= \mathrm{tr}[(\hat{\sigma}_i \otimes\hat{\sigma}_j)\hat\rho]$, 
and the Bloch vectors  $\mathbf{s}$ and $\mathbf{p}$
with elements $s_i=\mathrm{tr}[({\hat{\sigma}}_i\otimes{\hat{\sigma}}_0)\hat\rho]$ 
and $p_i=\mathrm{tr}[(\hat{\sigma}_0\otimes{\hat{\sigma}}_j)\hat\rho]$, respectively.
The matrices $\hat{\sigma}_i$ for $i=1,\,2,\,3$ are standard
Pauli matrices and $\hat{\sigma}_0$ is a single-qubit identity matrix. 
The invariants~\cite{Makhlin00} required to 
express negativity as described in
Refs.~\cite{Bart15a,Bart15b}
are \yyy{
\begin{equation}
\begin{aligned}
I_1= \det\hat\beta,\quad   I_2=\mathrm{tr}(\hat\beta^T\hat\beta),\quad  
I_3=\mathrm{tr}(\hat\beta^T\hat\beta)^2,\\
I_4=\mathbf{s}^2,\quad   I_5=[\mathbf{s}\hat\beta]^2,\quad  I_7 =\mathbf{p}^2,
I_8 =[\hat\beta\mathbf{p}]^2,\\   I_{12}=\mathbf{s}\hat\beta\mathbf{p},\quad   I_{14}= \varepsilon_{ijk}\varepsilon_{lmn}s_ip_l\beta_{jm}\beta_{kn},  
\end{aligned}\label{eq:MInv}
\end{equation}}
where  $\varepsilon_{ijk}$ is the Levi-Civita symbol.
Throughout this paper we use the Einstein summation convention.
Moreover, we will express the  double Levi-Civita symbol
in terms of Kroncker's delta symbols as shown, e.g., in Ref.~\cite{King06}. 
In the following sections we express these 9 invariants as the expected values 
of singlet-projections performed on multiple copies 
of a given two-qubit system. 

\section{Multicopy formulas for negativity and universal entanglement witness}

Here, we  further investigate the operational meaning 
of negativity and the universal entanglement witness
in the context of performing joint measurements on up to
four copies of a given two-qubit system in state $\hat{\rho}$. 
\yyy{This is a completely different approach than the one originally
based on consecutive parity measurements proposed in Ref.~\cite{Bart15a}.}
\yyy{As we demonstrate here, every} 
negativity-related invariant can be expressed as a 
function of positive valued measurements (projections) 
performed on multiple copies of the investigated two-qubit state.
These measurements are invariant under local unitary operations on $\hat{\rho}$.
The basic building block in our approach is  projection onto
singlet state, i.e.,  
$\hat P=(\hat{\sigma}_0\otimes\hat{\sigma}_0-\hat{\sigma}_i\otimes\hat{\sigma}_i)/4
\equiv|\Psi^-\rangle\langle\Psi^-|,$ where $i=1,\,2,\,3.$ 
\yyy{We construct multicopy observables for Makhlin's invariants
as explained on the following examples.}

\begin{figure}
\includegraphics[width=8.5cm]{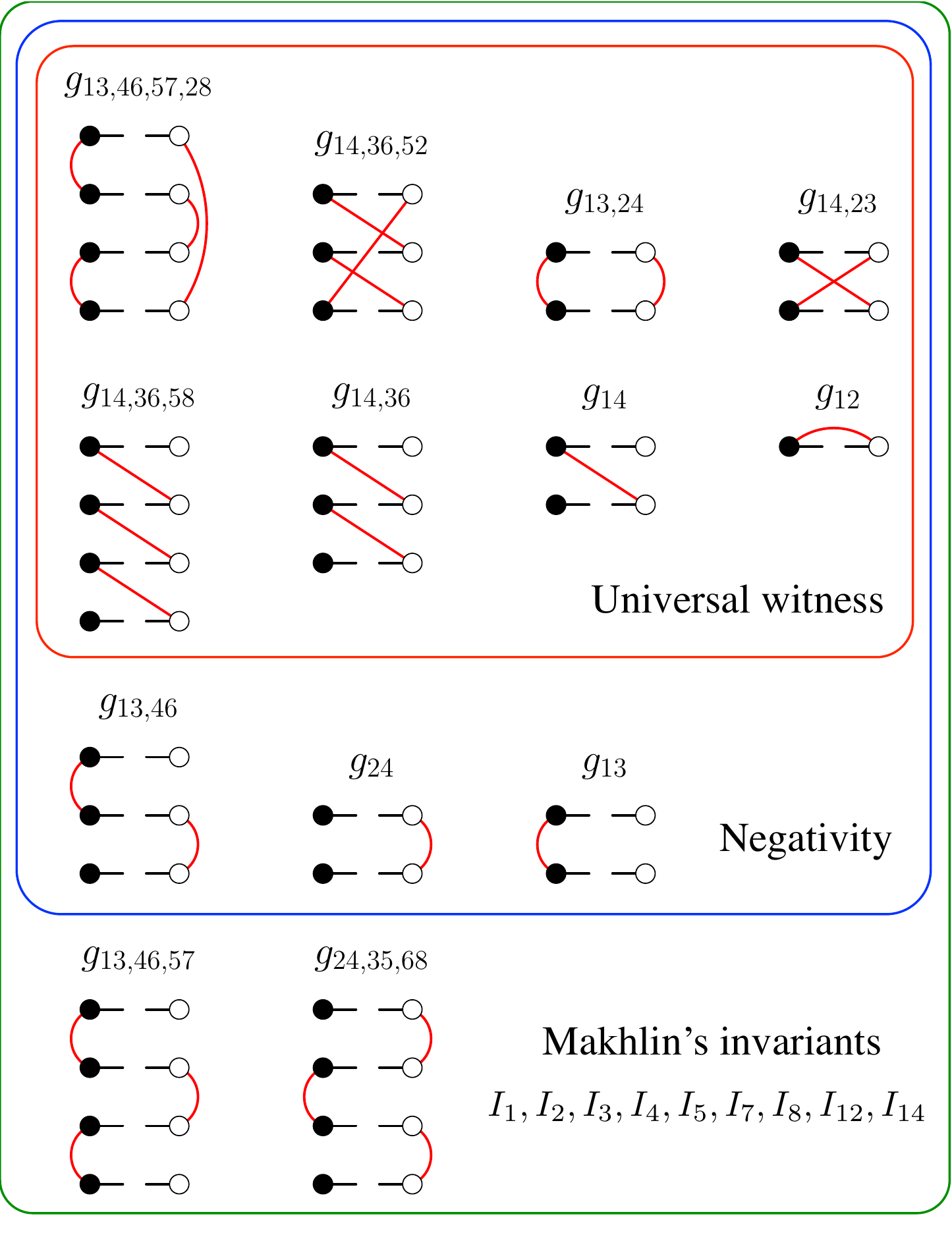}
\caption{\label{fig:1} (color online) The minimal 
set of singlet-projection-based observables needed to 
measure 9 negativity-related Makhlin's invariants
given in Eq.~(\ref{eq:MInv}), negativity defined in Eq.~(\ref{eq:neg}),
and universal entanglement witness from Eq.~(\ref{eq:drg}). 
Singlet projections are marked as solid curves, dashed lines 
combine subsystems (black and white discs) of the same copy of $\hat \rho$.}
\end{figure}

\yyy{As the first example let} us take 
$I_4=\mathbf{s}^2=\langle \hat{\sigma}^{(1)}_i\otimes \hat{\sigma}^{(2)}_0\rangle_{\hat\rho} 
\langle \hat{\sigma}^{(1)}_i \otimes\hat{\sigma}^{(2)}_0\rangle_{\hat \rho},$
where the subsystems are now
numbered and the observables are measured for a single copy 
of a system $\hat{\rho}$ and
$\langle \hat{\sigma}^{(1)}_i\otimes \hat{\sigma}^{(2)}_0\rangle_{\hat \rho} \equiv \mathrm{tr}[\hat{\sigma}^{(1)}_i\otimes \hat{\sigma}^{(2)}_0\hat\rho].$
To measure this invariant with an additional copy of the same system we
continue numbering the subsystems so that
the copies of the first and second subsystem are named
$3$ and $4,$ respectively. Hence, we have
$I_4=\langle \hat{\sigma}^{(1)}_i\otimes\hat{\sigma}^{(2)}_0
\otimes\hat{\sigma}_i^{(3)}\otimes\hat{\sigma}_0^{(4)} 
\rangle_{\hat\rho\otimes\hat \rho} =1-4\langle \hat{P}_{1,3}\rangle_{\hat \rho\otimes\hat \rho} \equiv 1-4g_{13},$
where the singlet projection is performed on the 
first and the third particle in the sequence.
Here, we introduce the notation ($g$ with the proper 
subscripts, see Fig.~\ref{fig:1}) 
that is used throughout the paper
to name the expected values of the
multi-copy observables. 

\yyy{In the second example let} 
us first expand $I_1$ in terms of the moments of matrix 
$\hat\beta$ as
\begin{equation}\label{eq:detb}
I_1 = \det\hat\beta = \tfrac{1}{6}[(\mathrm{tr}\hat\beta)^3 + 2 \mathrm{tr}\hat\beta^3 - 3 
\mathrm{tr}\hat\beta \mathrm{tr}\hat\beta^2]  
\end{equation}
We can express all these moments as 
\begin{eqnarray}
\mathrm{tr}\hat\beta &=& \langle \hat{\sigma}^{(1)}_i\otimes\hat{\sigma}^{(2)}_i \rangle_{\hat{\rho}},\nonumber \\
\mathrm{tr}\hat\beta^2 &=& \langle \hat{\sigma}^{(1)}_i\otimes\hat{\sigma}^{(2)}_j \otimes\hat{\sigma}^{(3)}_j\otimes\hat{\sigma}^{(4)}_i\rangle_{\hat{\rho}\otimes\hat{\rho}},\\
\mathrm{tr}\hat\beta^3 &=& \langle \hat{\sigma}^{(1)}_i\otimes\hat{\sigma}^{(2)}_j \otimes\hat{\sigma}^{(3)}_j\otimes\hat{\sigma}^{(4)}_k\otimes\hat{\sigma}^{(5)}_k\otimes\hat{\sigma}^{(6)}_i\rangle_{\hat{\rho}^{\otimes 3}},\nonumber 
\end{eqnarray}
where $\hat{\sigma}^{(a)}_i\otimes\hat{\sigma}^{(b)}_i=1-4\hat P_{a,b}.$ 
After some direct algebraic manipulations we are left with several equivalent
expected values. 
The equivalent terms are products of the same number of
$\hat P$ operators, and can be represented as $\langle\bigotimes_{(n,m)}\hat  P_{n,m} \rangle_{\hat\rho^{\otimes N/2}},$
where the tensor product $\bigotimes$ is taken over the relevant $N/2$ pairs of qubits $(m,n).$
We can find these terms by rearranging the order of copies of $\hat\rho$. 
Any two terms are equivalent, if we can find a natural number 
$k=1,2,3,4$ for which 
$\langle\bigotimes_{(n,m)} \hat P_{n,m} \rangle_{\hat\rho^{\otimes N/2}} =\langle\bigotimes_{(n,m)} \hat P_{n\oplus 2k,m\oplus 2k}\rangle_{\hat\rho^{\otimes N/2}},$ where
$\oplus$ stands for sum modulo the number of particles $N$, e.g.,
for $N=6$ we get $3\oplus 2=5$, $4\oplus2=6,$ $6\oplus 2=2$ etc.
After identifying equivalent terms in the analyzed expressions, 
the moments of $\hat\beta$ are given as
\begin{eqnarray}
\mathrm{tr}\hat\beta &=& 1-4 g_{12},\nonumber\\
\mathrm{tr}\hat\beta^2 &=& 1 - 8 g_{14} + 16 g_{14,23},\label{eq:beta}\\
\mathrm{tr}\hat\beta^3 &=& 1 - 12 g_{14} + 48 g_{14,36} - 64 g_{14,36,25}.\nonumber
\end{eqnarray}

\yyy{In the final  example  of  $I_{14}$ we first express the invariant} 
in terms of Kronecker's delta symbols
by means of an identity given, e.g., in Ref.~\cite{King06}.
This identity reads as
\begin{eqnarray}
\varepsilon_{ijk}\varepsilon_{lmn} &=& \delta_{il}\delta_{jm}\delta_{kn} +
\delta_{im}\delta_{jn}\delta_{kl} + \delta_{in}\delta_{jl}\delta_{km}\nonumber \\
&&-\delta_{il}\delta_{jn}\delta_{km} - \delta_{im}\delta_{jl}\delta_{kn}
-\delta_{in}\delta_{jm}\delta_{kl}.\quad
\end{eqnarray}
Now, we can rewrite $I_{14}= \varepsilon_{ijk}\varepsilon_{lmn}s_ip_l\beta_{jm}\beta_{kn}$ 
using the above mathematical identity and the methods introduced for $I_4$ and $I_1$ as
\begin{eqnarray}
I_{14} &=&  16 [g_{12}^2 (1 - 4 g_{14})  +2 g_{12} (  4 g_{14,36}-g_{14}) \nonumber\\
&&  - g_{14,23}  + 4 g_{14} g_{14,23}  + 2 g_{14,36} - 8 g_{14,36,58} ].
\end{eqnarray}

\yyy{We applied the techniques explained in the three presented 
examples to the relevant 9 invariants of Makhlin and after
calculations expressed them in terms of multi-copy measurements as}
\begin{eqnarray}
I_1 &=&  -\tfrac{8}{3}\lbrace g_{12} [g_{12} (4g_{12} -3)+6 (g_{14}-2 g_{14,23})] \nonumber\\
&& +3 g_{14,23} -6g_{14,36}+8g_{14,36,52}\rbrace,\nonumber\\
I_2 &=&  1 + 16 g_{13,24} - 4 (g_{13} + g_{24}),\nonumber\\
I_3 &=&  256 \left(g_{13}^2+4 g_{13,46}+g_{24}^2\right),\nonumber\\
&&-8 (g_{13}+g_{24})+256 g_{13,46,57,28}+1,\nonumber\\
I_4 &=&  1-4g_{13}, \label{eq:MinvG}\\
I_5 &=&  -4 g_{24} +32 g_{13,46} - 64 g_{13,46,57}+(1- 4 g_{13})^2,\nonumber\\
I_7 &=&  1-4g_{24},\nonumber\\
I_8 &=&  -4 g_{13} + 32 g_{13,46} - 64 g_{24,35,68} + (1 - 4 g_{24})^2,\nonumber\\
I_{12} &=& 1 + 16 g_{13,46} - 4 (g_{13} + g_{24}),\nonumber\\
I_{14} &=&  16 [g_{12}^2 (1 - 4 g_{14})  +2 g_{12} (  4 g_{14,36}-g_{14}) \nonumber\\
&&  - g_{14,23}  + 4 g_{14} g_{14,23}  + 2 g_{14,36} - 8 g_{14,36,58} ],\nonumber
\end{eqnarray}
where the relevant 13 terms  
$g_{12},$ $g_{13},$ $g_{14},$ $g_{24},$
$g_{13,24},$ $g_{13,46},$ $g_{14,23}$ $g_{14,36},$
$g_{14,36,52},$  $g_{13,46,57},$  $g_{24,35,68},$ 
 $g_{13,46,57,28},$  $g_{14,36,58},$ 
are defined as expected values of
projections on multiple singlet states as shown in Fig.~\ref{fig:1}. 
This result allows us to study the state-dependent parameters 
\begin{eqnarray}
a_0&=& -16 [g_{12}^3 + 2g_{14,36,52}+ \nonumber \\
&& 3( g_{13,24}^2  -  g_{12}^2 g_{14} -  g_{12} g_{14,23} + g_{14} g_{14,23} )\nonumber\\
&&   - 6 (g_{13,46,57,28}-  g_{12} g_{14,36}  +    g_{14,36,58})]\nonumber  \\
a_1&=& 24[g_{12}^2 - g_{14,23} - g_{13,24}\\
&&+ 2(g_{13,46} - g_{12}g_{14}+g_{14,36})] \nonumber\\
&&    - 32(g_{12}^3 -3g_{12}g_{14,23}  +2g_{14,36,52}),\nonumber\\
a_2&=&12(g_{13} - 2g_{13,24} + g_{24}), \nonumber
\end{eqnarray}
\yyy{needed to calculate the negativity with Eq.~(\ref{eq:neg})}
as functions of the multicopy observables. 
It turns out that these coefficients are expressed with 11 terms, i.e.,
$g_{12},$  $g_{13},$ $ g_{14},$  $ g_{24},$ 
$g_{13,24},$ $g_{13,46},$    $g_{14,23},$ $g_{14,36},$ 
$g_{14,36,52},$  
$g_{13,46,57,28},$   $ g_{14,36,58}.$
The universal entanglement witness 
in terms of singlet projections can be expressed as $\det\hat\rho^\Gamma = a_0/48,$ where
a given two-qubit state is entangled if and only if $\det\hat\rho^\Gamma <0.$
However, to measure negativity one needs to know
the values of $a_n$ for $n=0,1,2$.
Note, that to witness  entanglement it is enough to measure
a smaller set of observables than for
negativity. This set has 8 elements and it 
does not include the $g_{13}, g_{24}, g_{13,46}$ measurements.
Thus, these measurements contain the extra information
that is needed to quantify the entanglement instead 
of simply detecting it. Our analysis of the solutions to the quartic Eq.~(\ref{eq:neg})
with the help of a computer algebra system did not reveal
any further reductions in the number of measurements
needed to estimate the negativity.

\section{Optical implementation of minimal set of multicopy projections}

\begin{figure}
\includegraphics[width=6cm]{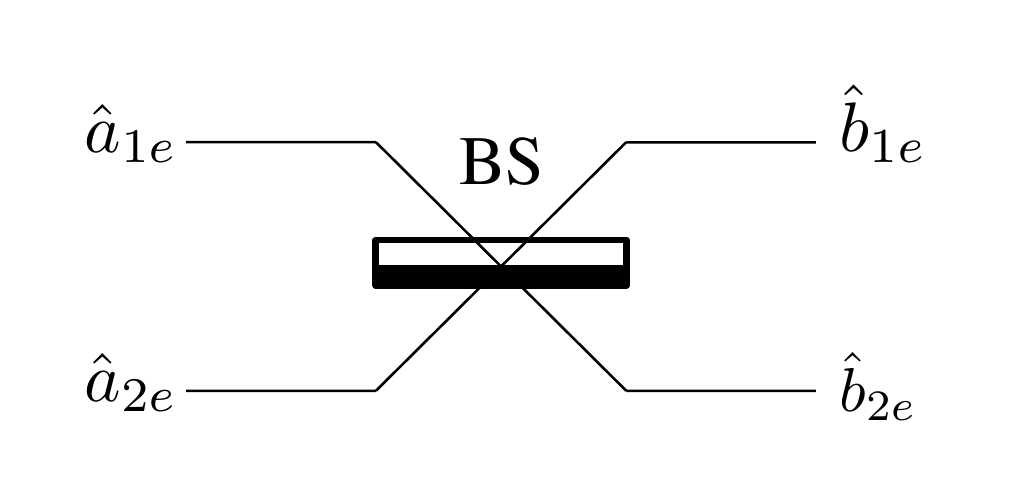}
\caption{\label{fig:3}  The 50:50 beam-splitter (BS) transforms the input  annihilation operators $\hat a_{1e}$
and $\hat a_{2e}$ into output  annihilation operators $\hat b_{1e}$ and $\hat b_{2e}$ 
according to $\hat a_{1e}=(\hat b_{1e}+\hat b_{2e})/\sqrt{2}$ and $\hat a_{2e}=(\hat b_{1e}-\hat b_{2e})/\sqrt{2},$ 
where  $e=H,V$ represents two orthogonal polarization modes (see, e.g., ~\cite{KokBook}).}
\end{figure}

The singlet projection $\hat P$  is frequently applied to investigate the
quantum properties of polarization-encoded
two-qubit states~\cite{Bovino05,Jin2012,Bartkiewicz13discord,Bartkiewicz13fidelity,Miran14,Bartkiewicz16fef}.
In this case, density matrix $\hat \rho$ describes 
a pair of polarization-encoded qubits with Pauli matrices 
$\hat{\sigma}_1 = |D,D\rangle \langle D,D| - |A,A\rangle \langle A,A|,$
$\hat{\sigma}_2 = |L,L\rangle \langle L,L| - |R,R\rangle \langle R,R|,$ and
$\hat{\sigma}_3 = |H,H\rangle \langle H,H| - |V,V\rangle \langle V,V|,$ which
are expressed in terms of diagonal ($|D\rangle$), anti-diagonal ($|A\rangle$),
left-circular ($|L\rangle$), right-circular ($|R\rangle$), horizontal ($|H\rangle$),
and vertical ($|V\rangle$) polarization states.
The singlet projection $\hat P$ can be implemented 
by measuring the anti-coalescence rate of photons that interfered 
on a balanced beam splitter (BS). 
Any two-qubit state can be expressed in a basis of
the four following maximally-entangled states
\begin{eqnarray}
 \ket{\Psi^{\pm}} &=& \frac{1}{\sqrt{2}}(\ket{H,V}\pm\ket{V,H}),\nonumber\\
  \ket{\Phi^{\pm}} &=& \frac{1}{\sqrt{2}}(\ket{H,H}\pm\ket{V,V}).
\end{eqnarray}
We can express these two-photon states in terms of the creation 
operators $\hat{a}_{1e}$ and $\hat{a}_{2e}$ for polarizations
$e=H,V$ (see Fig.~\ref{fig:3}), \yyy{where, e.g.,
$\ket{V,H}=\hat{a}_{1V}^{\dagger}\hat{a}_{2H}^{\dagger}\ket{0,0}$ and
$\ket{0}$ is the vacuum.} 
Next, the states are  transformed by the BS (see Fig.~\ref{fig:3})
as follows\yyy{
\begin{eqnarray}
  U_{\rm BS} \ket{\Psi^{-}} &=& -\ket{\Psi^{-}},\nonumber
\\
  U_{\rm BS} \ket{\Psi^{+}} &=& \frac{1}{\sqrt{2}}(\hat{a}_{1V}^{\dagger}\hat{a}_{1H}^{\dagger}-\hat{a}_{2V}^{\dagger}\hat{a}_{2H}^{\dagger})\ket{0,0},
\\
  U_{\rm BS} \ket{\Phi^{\pm}} &=&
  \frac{1}{2\sqrt{2}}(\hat{a}_{1H}^{\dagger2}-\hat{a}_{2H}^{\dagger2}  \pm \hat{a}_{1V}^{\dagger2}  \mp\hat{a}_{2V}^{\dagger2})\ket{0,0}.\nonumber
\end{eqnarray} }
Thus, observing anti-coalescence is equivalent to performing 
a singlet projection. We will use this well known fact~\cite{HOM,Pan12}
to design specialized interferometers 
to detect and measure the entanglement
of an arbitrary two-qubit state. 

\begin{figure*}
\includegraphics[width=16cm]{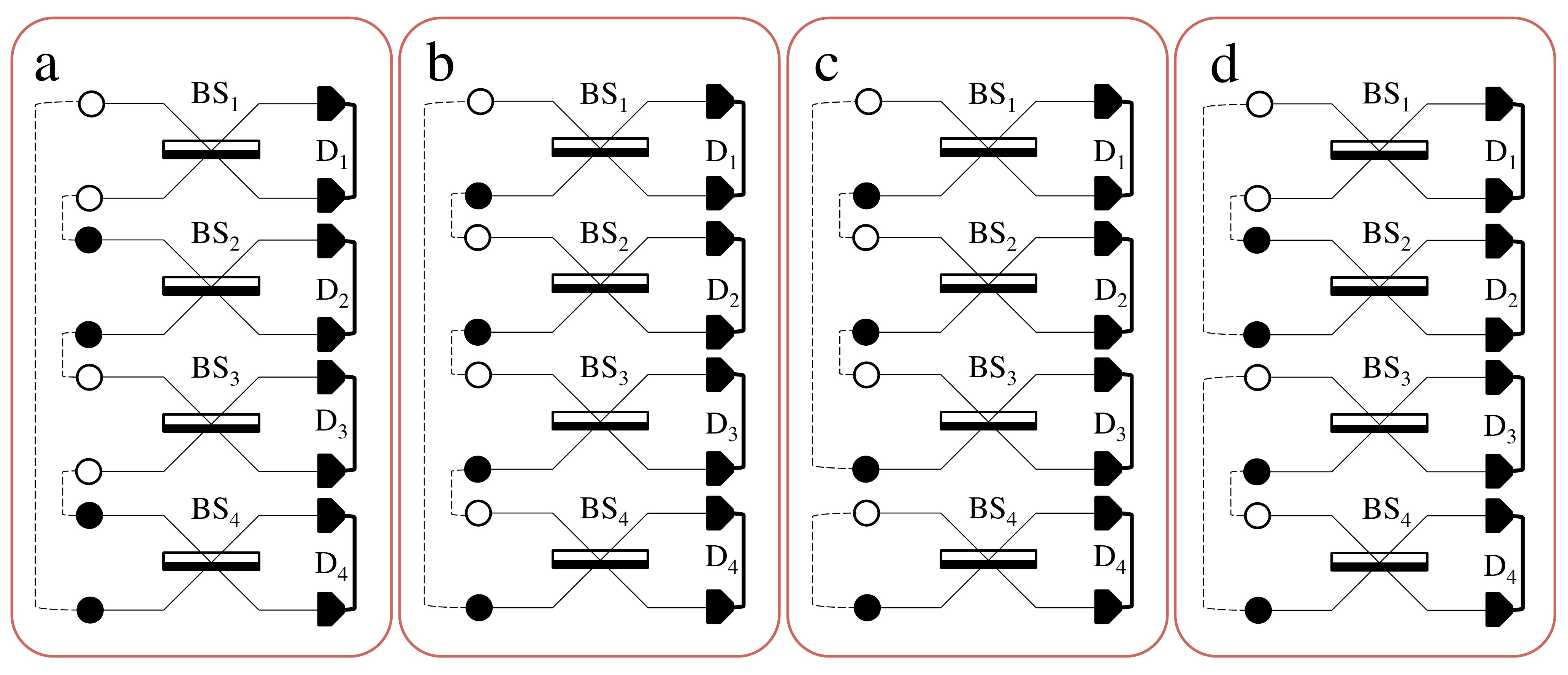}
\caption{\label{fig:2} (color online) 
Interferometric configurations for measuring all independent  observables 
from Fig.~\ref{fig:1}
with four copies of polarization-encoded $\hat\rho$. 
In configuration (b) our the interferometer can
additionally measure $g_{14,36,58,72},$
which can be expressed as a function
of other quantities by means of the Cayley–Hamilton theorem [see Eq.~(\ref{eq:gHC})].
Subsystems of a single copy
are depicted as black and white discs connected with dashed lines.
Photons interfere on beam splitters BS$_n$ for $n=1,2,3,4$ and their
coalescence or anti-coalescence is detected by detector modules D$_n$
for $n=1,2,3,4$ (see, e.g., Ref.~\cite{Bovino05}). 
For detailed analysis of all the possible detection events
see Tab.~\ref{tab:1}.}
\end{figure*}

The measurements that can be used to determine the 9 relevant 
Makhlin's invariants can be grouped into 6 sets.
The first two sets of measurements are
$S_1=\{g_{13,46,57,28},g_{13,46,57},g_{24,35,68},g_{13,46},g_{13},g_{24}\}$
and 
$S_2=\{g_{14,36,58},g_{14,36,58},g_{14,36},g_{14}\}.$
All the elements in these sets can be measured
with interferometers that measure $g_{13,46,57,28}$ 
or $g_{14,36,58}$ on four copies of a given state. 
A proper analysis of the coincidence counts
provides values of the remaining less complex
measurements from this set (see Tab.~\ref{tab:1}).
The next measurement set is $S_3=\{g_{14,36,52}, g_{14,36}, g_{14}\},$
where all the relevant outcomes can be obtained
with an interferometer designed to measure $g_{14,36,52}$
on three copies of $\hat \rho$. The last three measurement sets are 
$S_4=\{g_{13,24},g_{13},g_{24}\}$, $S_5=\{g_{14,23}, g_{14}\},$ 
and $S_6=\{g_{12}\}$ which can be measured with three interferometers 
operating with two or one copy of $\hat \rho.$ 
However, to measure all the above-listed quantities 
with four copies of $\hat \rho$ we need no more than 
four experimental configurations in total.
These three configurations measure (a) $S_1$, (b) $S_2,$ (c)  $S_3$ and $S_6$, (d) $S_4$ and $S_5$
are shown in the respective panels of Fig.~\ref{fig:2}. 
Note, that some measurements (e.g., $g_{14},$ $g_{13},$ and $g_{24},$) 
are performed in more than one configuration (see Tab.~\ref{tab:1}). 

In configuration (b) the interferometer
measures observable $g_{14,36,58,72},$ which appears in
the following expression for the fourth moment of $\hat\beta,$ i.e.,
\begin{eqnarray}
\mathrm{tr}\hat\beta^4 &=& 1 - 16 g_{14} + 32(2 g_{14,36}+g_{14}^2)\nonumber\\
&& + 256 ( g_{14,36,58,72}-g_{14,36,58}).
\end{eqnarray}
Thus, we have
\begin{eqnarray}
g_{14,36,58,72}&=& \tfrac{1}{256}[\mathrm{tr}\hat\beta^4 - 1 + 16 g_{14} 
\nonumber\\
&&-32(2 g_{14,36}+g_{14}^2)]+ g_{14,36,58},\label{eq:gHC}
\end{eqnarray}
where $\mathrm{tr}\hat\beta^4$ is calculated using
the Cayley-Hamilton theorem (see, e.g., Ref.~\cite{Jing2015}) for $\hat\beta,$  i.e.,
\begin{equation}
\mathrm{tr}\hat \beta^4 =   \mathrm{tr}\hat \beta 
- \tfrac{1}{2}\mathrm{tr}\hat \beta^2 (\mathrm{tr}^2 \hat \beta-\mathrm{tr}\hat \beta^2 ) + \mathrm{tr}\hat \beta \det\hat\beta,
\end{equation}
where the moments  $\mathrm{tr}\hat \beta^n$ for $n=1,2,3$ 
are defined in Eq.~(\ref{eq:beta}) and the determinant 
$\det\hat\beta$ in Eq.~(\ref{eq:detb}) or Eq.~(\ref{eq:MinvG}). 
Thus, observable $g_{14,36,58,72}$ can be expressed
using the observables listed in Fig.~\ref{fig:1}.

\begin{table*}
\caption{\label{tab:1} {
Interpretation of detection events of the interferometers shown in Fig.~\ref{fig:2}.
Each couple of detectors D$_n$ for $n=1,2,3,4$ detects coalesce or
anti-coalescence for a pair of impinging photons.
The accumulated counts of (anti-) coalescence events can be 
grouped into $c$ coalescence or $s=a+c$ sum of coalescence ($c$)
and anti-coalescence ($a$). Thus, the total number of all detection events
is $Z$. Depending on the measured quantity, one can choose
the required detection events in accord with Fig.~\ref{fig:1}.}}
\begin{ruledtabular}
\begin{tabular}{ccccccccc}
D$_1$ & D$_2$ & D$_3$ & D$_4$ & Fig. ~\ref{fig:2}a & Fig. ~\ref{fig:2}b & Fig. ~\ref{fig:2}c & Fig. ~\ref{fig:2}d \\
\hline
s & s & s & s & $Z$ & $Z$ & $Z$ & $Z$ \\
s & s & s & a & $Zg_{24}$ & $Zg_{14}$&$Zg_{12}$ & $Zg_{14}$ \\
s & s & a & s & $Zg_{13}$ & $Zg_{14}$& $Zg_{14}$ & $Zg_{14}$\\
s & s & a & a & $Zg_{13,46}$ & $Zg_{14,36}$ & $Zg_{14}g_{12}$& $Zg_{14,23}$\\
s & a & s & s & $Zg_{24}$ & $Zg_{14}$& $Zg_{14}$ & $Zg_{24}$\\
s & a & s & a & $Zg_{24}^2$ & $Zg_{14}^2$ & $Zg_{14}g_{12}$& $Zg_{24}g_{14}$\\
s & a & a & s & $Zg_{13,46}$ & $Zg_{14,36}$& $Zg_{14,36}$& $Zg_{24}g_{14}$\\
s & a & a & a & $Zg_{24,35,68}$ & $Zg_{14,36,58}$ &$Zg_{14,36}g_{12}$& $Zg_{24}g_{14,23}$\\
a & s & s & s & $Zg_{13}$ & $Zg_{14}$ &$Zg_{14}$ & $Zg_{13}$\\
a & s & s & a & $Zg_{13,46}$ & $Zg_{14,36}$ &$Zg_{14}g_{12}$ & $Zg_{13}g_{14}$\\
a & s & a & s & $Zg_{13}^2$ &  $Zg_{14}^2$ &$Zg_{14,36}$ & $Zg_{13}g_{14}$\\
a & s & a & a & $Zg_{13,46,57}$ & $Zg_{14,36,58}$ &$Zg_{14,36}g_{12}$ & $Zg_{13}g_{14,23}$\\
a & a & s & s & $Zg_{13,46}$ & $Zg_{14,36}$ &$Zg_{14,36}$ & $Zg_{13,24}$\\
a & a & s & a & $Zg_{24,35,68}$ & $Zg_{14,36,58}$ &$Zg_{14,36}g_{12}$  & $Zg_{13,24}$\\
a & a & a & s & $Zg_{13,46,57}$ &$Zg_{14,36,58}$  &$Zg_{14,36,52}$ & $Zg_{13,24}g_{14}$\\
a & a & a & a & $Zg_{13,46,57,28}$  & $Zg_{14,36,58,72}$  & $Zg_{14,36,52}g_{12}$ & $Zg_{13,24}g_{14,23}$ \\
\end{tabular}
\end{ruledtabular}
\end{table*}

\section{Conclusions}

\yyy{Finding a minimal set of 13 interferometric 
quantities for expressing the relevant 9 Makhlin's
invariants (11 for negativity, and  8 for detecting 
entanglement of a given two-qubit state) 
is the main results of this paper. 
It explicitly proves that one has to 
perform more measurements to reconstruct the state (i.e., 15 measurements) 
than, e.g., to measure the negativity (i.e., 11 measurements). 
In contrast to the previous works~\cite{Carteret05,Bart15a,Bart15b},
here we explicitly demonstrated that all the necessary 
data for detecting or quantifying the entanglement
can be directly measured without collecting irrelevant 
information about the state. 
This was not apparent before, because the previously proposed 
measurement schemes were designed for measuring moments 
of a given partially transposed density matrix~\cite{Carteret05,Bart15a,Bart15b} 
and required ignoring some detection events or output modes,
or using ancillary entangled states. 
The interferometers shown in Fig.~\ref{fig:3}
measure only the functions of 13 observables 
depicted in Fig.~\ref{fig:1} and they
cannot be further simplified without loosing the ability to measure
the entanglement or the relevant 9 Makhlin's invariants.
Measuring local invariants with linear optics
requires collecting less data than performing 
a complete quantum state tomography, which 
for a two-qubit state requires 15 measurements. 
Hence, we also demonstrated that local invariants are both 
useful theoretical concept 
for designing specialized quantum interferometers and that
their direct measurement within the framework of linear optics 
does not require performing complete quantum state tomography.}

The described set of 11 observables is the minimal set 
of measurements needed
to determine the value of the negativity. 
Because one cannot express the basic measurements as functions of each other,
the presented set seems impossible to reduce further. 
Moreover, any attempt to discard
some of the measurements will change the values of
parameters $a_n$ for $n=0,1,..2$ in the characteristic equation,
thus, the value of $N$ calculated from Eq.~(\ref{eq:neg}). 
In contrast to the results presented in Ref.~\cite{Bart15a,Carteret05},
we do not need ancillary qubits \yyy{and we use information
from all output modes.}

Our results provide a new perspective on the phenomenon
of quantum entanglement \yyy{in terms of entanglement cost
under PPT operations.}  We demonstrated in Figs.~\ref{fig:1} and \ref{fig:3}
that two-qubit entanglement can be fully described using two-photon
interference events between subsystems of at most four copies
of a given state.  As explicitly shown in Tab.~\ref{tab:1}, 
our approach gives us only the information needed 
to measure negativity, universal entanglement witness, and
the relevant Makhlin's invariants.
All the measured information 
can be interpreted in terms of the minimal
set of observables depicted in Fig.~\ref{fig:1}.
This approach only requires using 
beam splitters and photon detectors, i.e., the basic 
building blocks of quantum information processing
within the framework of linear optics~\cite{KokBook}.  
However, singlet projections on multilevel systems
can be also implemented in, e.g., solid state systems ~\cite{Tanaka14}.

The presented general approach can be also used for measuring
a different type of quantum correlations than quantum entanglement
~\cite{Modi12}, i.e. quantum discord. This type of quantum correlations
 is hard to compute (NP-complete) as shown in Ref.~\cite{Huang14}.
Note, that measuring or detecting geometric quantum discord 
could require more complex measurements than in the case of
entanglement, as described in Refs.~\cite{Jin2012,Bartkiewicz13discord}. 

\yyy{One of the open problems related to the topic of this paper
is the degree of complexity of analogous interferometers used for 
entanglement measures other than negativity. 
By studying this problem one could categorize 
the entanglement measures operationally with respect to the amount of experimental 
effort required to measure them. We expect that this would also 
give us some intuition about the experimental differences between 
the particular entanglement measures like, e.g., concurrence and
negativity, whose definitions are often too abstract to directly compare.}

\begin{acknowledgments}
We thank Adam Miranowicz for the stimulating discussions. 
K.B. and K.L. acknowledge the financial support by the Czech Science
Foundation under the project 
No. 16-10042Y; and the financial support of the 
Polish National Science Centre under the grants No. DEC-2013/11/D/ST2/02638 
(Sec. IV) and  No. DEC-2015/19/B/ST2/01999 (Sec. III); and the project No. LO1305 of the Ministry of Education, 
Youth and Sports of the Czech Republic.

\end{acknowledgments}

\end{document}